\documentclass[runningheads]{llncs}
\usepackage{amsfonts}       % blackboard math symbols
\usepackage{nicefrac}       % compact symbols for 1/2, etc.
\usepackage{microtype}      % microtypography
\usepackage{textcomp}
\usepackage{graphicx}
\usepackage{amsmath}
\usepackage{amssymb}
\usepackage{subcaption}
\usepackage{dirtytalk}
\begin{document}
\title{Minimizing Acquisition Maximizing Inference - A demonstration on print error detection}
\titlerunning{Privacy preserving print error detection}
\author{Suyash Shandilya}

\institute{Defence Research and Development Organisation\\
New Delhi - 110054, India\\
\email{su.sh2396@gmail.com}\\
}
\maketitle              % typeset the header of the contribution
\begin{abstract}
Is it possible to detect a feature in an image without ever looking at it? Images are known to have sparser representation in Wavelets and other similar transforms. Compressed Sensing is a technique which proposes simultaneous acquisition and compression of any signal by taking very few random linear measurements $(M)$. The quality of reconstruction directly relates with $M$, which should be above a certain threshold for a reliable recovery. Since these measurements can non-adaptively reconstruct the signal to a faithful extent using purely analytical methods like Basis Pursuit, Matching Pursuit, Iterative thresholding, etc., we can be assured that these compressed samples contain enough information about any relevant macro-level feature contained in the (image) signal. Thus if we choose to deliberately acquire an even lower number of measurements - in order to thwart the possibility of a comprehensible reconstruction, but high enough to infer whether a relevant feature exists in an image - we can achieve accurate image classification while preserving its privacy. Through the print error detection problem, it is demonstrated that such a novel system can be implemented in practise.

\end{abstract}
\section{Introduction}

\subsection{The Print error detection problem}
\label{sec:ped}

Printing errors are a very pervasive problem in large scale printing mills. Generally any conspicuously erroneous material is reprinted while innocuous ones are left as it is. But when the data being printed is sensitive (mathematical equations, numbers, encryption keys, etc.) or if the user is cautious about aesthetics, a single ambiguity in reading can impair an entire document. The cost of reprinting is usually infinitesimal in comparison to the jeopardy caused by an unusable copy. Thus a system to proofread every sheet of every document is imperative in such cases.
In general, this can be easily automated by training standard classifiers over a large dataset. The critical issue is, many users do not want their classified data to be digitised. From cyberattacks to system compatibility, there are many reasons why organisations avoid digitisation of sensitive material. Thus proofreading in such cases needs to be done manually. This is expensive in terms of human resources, time, et cetera. This seemingly makes the problem impossible to automate unless there could be a system which can detect these errors without digitally 'looking' at the data.

\subsection{Compressed Sensing}
\label{sec:cs}
Compressed sensing has been a prolific research topic over the last few years. Basics of compressed sensing can be found in studies by Candes,Romberg, Tao \cite{candromtao} and Donoho \cite{donoho}. Readers may refer to \cite{cstut} as an excellent introductory tutorial. A very brief introduction is as follows.

A typical signal acquisition process conventionally begins with sampling the signal in spatial or time domain at a frequency much higher than the Nyquist sampling rate \cite{samplingtheorem}. Most of the signals (specially images) have a lot of redundancy at this stage. As a practical and pertinent example, a 10 megapixel RGB camera with a bitdepth of 16 bits per pixel should produce a 20 MB image. But the JPEG file format compresses the image to about a tenth of its size, if not more. For almost all practical purposes, this compressed file is as good as the raw image with the added convenience of portability. This depicts an innate inefficiency in the system wherein we first deploy expensive hardware to acquire data at a high sampling frequency and subsequently discard most of the samples.

Compressed sensing proposes that if the signal is sparse in some domain (natural images are sparse in wavelets, DCT) then a sufficient number of incoherent linear non-adaptive measurements of the signal can be used to recover the signal as accurately as done by any conventional sampling system. This simplifies the acquisition process and greatly reduces the hardware complexity. The reconstruction however, is more computationally expensive.

Essentially, we are simply trying to obtain the sparsest possible representation of our signal. This can be mathematically expressed as:

\begin{equation*}
\label{eq:loprob}
    \begin{aligned}
        & \underset{x}{\text{minimize}}
        & & ||x||_0 \\
        & \text{subject to}
        & & y = \phi x
    \end{aligned}
\end{equation*}

Where $y \in \mathbb{R}^M$ is the obtained compressed sample vector, $x \in  \mathbb{R}^N$ is the original \textbf{sparse} signal and $\phi \in  \mathbb{R}^{M\times N}$ is called as the \textit{sensing matrix} ($M<<N$). Here $||.||_0$ denotes the $l_0$-norm which is equal to the number of non-zero components in a vector. This is known to be an NP-hard problem \cite{donoho}. The real genius of compressive sensing lies in proving that under certain conditions of noise and sparsity, this problem is equivalent to its convex relaxation: 

\begin{equation*}
\label{eq:l1prob}
    \begin{aligned}
        & \underset{x}{\text{minimize}}
        & & ||x||_1 \\
        & \text{subject to}
        & & y = \phi x
    \end{aligned}
\end{equation*}

This problem has been termed as \textit{Basis Pursuit}. More practical version of the problems like \textit{Basis Pursuit Denoising} (BPDN) and LASSO are also popular in literature. Readers are advised to refer to \cite{pareto} for more details on its variants.

Often $x$ is not sparse in the original domain but has a sparse representation in a transformed domain like wavelets, etc. In that case, $x$ is replaced with $\psi'x$ where $\psi$ is the transformation matrix of the domain in which $x$ is sparse.

The sensing matrix $\phi$ is very crucial to the concept here. It describes how the linear measurements are obtained from $x$. The number of columns in $\phi$ should be equal to the length of the signal $x(=N)$ while the number of rows of $\phi$ determine the number of compressed samples ($M$) the user wants. Higher the $M$ better the reconstruction accuracy. Two important properties are associated with the sensing matrix: \textit{Incoherence} and \textit{Restricted Isometry Property} (RIP). Simply stated, they ensure that the sensing matrix takes incoherent (well-spread) measurements, and disparate signals should not map to similar measurement values. This sensing matrix is expected to be maximally incoherent with the transformation matrix $\psi$. It was found that random matrices like Gaussian, Bernoulli matrix satisfy both the aforementioned properties with a very high probability. Here binary Gaussian matrix has been used for reasons better described in \ref{sec:spc}.

Greedy algorithms like Orthogonal Matching pursuit \cite{omp} and many of its variants like ROMP \cite{romp}, StOMP \cite{stomp}, CoSaMP \cite{cosamp}, are also used to recover the sparse signal. They are much faster than Basis Pursuit methods but have poor theoretical guarantees. Since the objective of the paper is to ensure minimal signal recovery in the worst possible case, I adhered to evaluating the recovered signals via Basis Pursuit only.

\subsection{Single Pixel Camera and Smashed Filters}
\label{sec:spc}

The first practical implementation of compressed sensing was demonstrated by Duarte, et al. when they created a single pixel camera (SPC) \cite{spc}. They used a Digital Micromirror Device (DMD) which is an array of small mirrors which can be digitally aligned along two binary states (-10\textdegree or +10\textdegree from the axis). It is set in a way that one of the states aligns with the single photon detector while other reflects away from it. This essentially simulates a binary matrix. Since the system proposed in this work proposes this hardware, hence only binary matrix was chosen for all the experimentation in this paper. The reader may refer to \cite{spc} for a more elaborate understanding of the working of the model device.
In a similar work, Davenport et al. introduced the concept of \textit{smashed filters} \cite{smash}. Therein the authors posit that operations like target classification can be performed on the compressed samples themselves by applying maximum likelihood classifier. This has been one of the primary motivation for this work. More on this in Section \ref{sec:prevwork}.

% Smashed Filter: Maximum Likelihood Estimator. Mine is not a smashed filter. That's just a motivation.
\subsection{Dataset}
\label{sec:dataset}

\begin{figure}[th]
     \centering
     \begin{subfigure}[th]{0.3\textwidth}
         \centering
         \includegraphics[scale = 1]{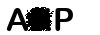}
         \caption{Text: AXP. Blot errors caused by ink splats.}
         \label{fig:be}
     \end{subfigure}
     \hfill
     \begin{subfigure}[th]{0.3\textwidth}
         \centering
         \includegraphics[scale = 1]{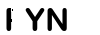}
         \caption{Text: RYN. Line error caused by printhead drags.}
         \label{fig:lpe}
     \end{subfigure}
     \hfill
     \begin{subfigure}[th]{0.3\textwidth}
         \centering
         \includegraphics[scale = 1]{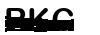}
         \caption{Text: PKC. Line errors caused by printhead slips.}
         \label{fig:lse}
     \end{subfigure}
        \caption{\small The 3 major type of print errors found in printing press. Other types of errors in most part are subsumed in either of these categories. Note that these are simulations of zoomed-in scans of text segments. A typical sheet may contain more than 500 such words. In terms of pixel ratio, each word shown here is about 0.1\% of the full sheet.}
        \label{fig:pe}
\end{figure}

No pre-existing digital repertoire of common print errors could be found. So the experiments here have been conducted on simulated print errors. They have broadly been binned into three major types of errors as shown in Fig. \ref{fig:pe}. These artifacts have been chosen to be just big enough to cause an ambiguity in the text.
The text images were generated using the Python Imaging Library (PIL). All possible 3 letter words were made as grayscale images of pixel dimension $35 \times 100$. Arial bold font was used. They were labelled as ‘good’ images. The three types of errors were distributed equally on the entire dataset of $17576 (= 26 \times 26 \times 26)$ images. These were labelled as bad images.

\subsection{Paper Layout}
\label{sec:layout}

% There was a good amount of skepticism between the conception and testing of this concept. As a fail-safe another application was conceived - It would be proposed as a safer way of obtaining any image analysis as only a user with the exact key will be able to reconstruct the meaningful image. Fortunately (and mildly surprisingly), that became a secondary application as even extreme levels of compression gave remarkable results.

Section \ref{sec:prevwork} discusses similar works which have preceded this paper. Sections \ref{sec:enc},\ref{sec:dec} explain how the data was compressed and reconstructed (only for testing) respectively. Section \ref{sec:results} discusses the chosen models for classification and compares their performance. Section \ref{sec:futwork} describes other groundbreaking applications where this concept should be tested.
\subsection{A Note on Terminology}
\label{sec:termino}
The words compression and encryption are used interchangeably throughout the text.
To be technically precise, the concept is to draw inference from a heavily (lossy) compressed data. But the novel idea of this work is the application of such a system to data privacy. Thus - although not technically accurate - a more application oriented description of the work would be the idea to draw inference from a severely \textit{encrypted} data.
\begin{itemize}
    \item Key $\equiv$ sensing matrix $(\phi)$.
    \item Encryption $\equiv$ compression $(x \to \phi x = y)$
    \item Decryption $\equiv$ Reconstruction $(\phi x = y \to x)$
\end{itemize}
A similar terminology has also been described in \cite{ppcs} (II-C).

\section{Previous Work}
\label{sec:prevwork}

Probably the first work to demonstrate Compressive Sensing's capabilities in target classification was the Single Pixel Camera developed by Davenport et al. \cite{smash}. Their results bespeak that effective classification could be achieved from compressed samples without any need of reconstruction. Therein the authors introduce smashed filters which uses maximum likelihood for classification (MLE)\footnote{This is a major difference between this work and \cite{smash}. In the case of AWGN, the generalised MLE is equivalent to \textit{nearest neighbour} classification. We tried using the same in our problem only to find it sub-par in comparision to other classifiers like polynomial discriminants and SVMs.}. The concept is demonstrated on images of toy tanker, school bus and truck and the accuracy results are caculated using leave-one-out validation\footnote{ All the results mentioned in this work are holdout validation results (the testing dataset was never used anywhere in evaluating the discriminant) unless stated otherwise.}. Since there were not privacy concern in their objective, hence there is no analysis in that regard. In fact there have been several other demonstrations of the application of smashed fiters in different domains (\cite{irtrack,ttw,jrom} to cite a few) but none seem to have designed or evaluated a system with privacy concerns as priority.\\
Compressed sensing has been extensively studied from an encryption system standpoint. \cite{crypt1} was the first work where it was analysed as a symmetric encryption system. The authors prove that it does not provide information theoretic secrecy but argue that it is computationally secure, as long as the sensing matrix is used only once. Authors of \cite{ppcs} assert its effectiveness as an obfuscation layer in a cryptosystem. In \cite{crypt18}, the authors prove the computational security of CS against a systematic search of the sensing matrix, even when the signal sparsity is known. A practical CS system with two confidentiality levels based on Bernoulli sensing matrices has been proposed and analyzed in \cite{crypt19}, \cite{crypt20} and its security has been additionally investigated in \cite{crypt21}. In all such analyses the sensing matrix is considered to be available for the intended receiver and the data needs to be recovered reliably. It should be noted that both these assumptions are inconsequential here as we intentionally \textit{crumble} the data to ensure its unusability.\\
Another noteworthy contribution has been made by Zhou \cite{pprs} where the authors use compressed sensing as a matrix masking \cite{matrixmasking} tool on an entire statistical dataset to achieve \say{privacy preserving regression analysis}. To the best of my knowledge, this is the first work to demonstrate the simultaneous efficacy of compressed sensing in ensuring image security and high accuracy in image classification.

% \section{Encryption and Decryption}
\section{Data Encryption}
\label{sec:enc}

\begin{figure}[h]
    \centering
    \includegraphics[scale = 0.45]{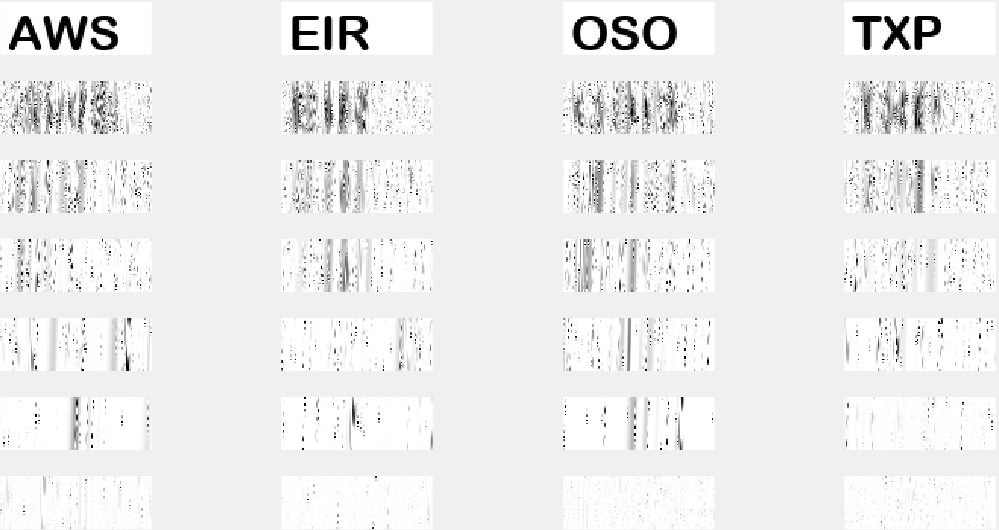}
    \caption{The top row corresponds to the original image. The second row corresponds to reconstruction using 500 compressed samples ($M = 500$, \textbf{only shown for reference, not used for classification}) where the characters are faintly visible. The subsequent rows correspond to $M = 200,100,50,20,10$ respectively.}
    \label{fig:recresults}
\end{figure}

Encrypted data ($y \in \mathbb{R}^M$) is achieved by the multiplication of a binary sensing matrix ($\phi \in \{0,1\}^{M\times N}$) with the original data ($x \in \mathbb{R}^N$). This is a form of lossy encryption as $M << N$, thus the data is not likely to be fully recovered. Nevertheless, any level of accurate decryption requires $\phi$ to be known exactly, thus the sensing matrix is essentially the \textit{key} for this encryption here.
For a successful recovery, we need to choose a basis in which the data is represented sparsely enough. Daubechies – 10 wavelets \cite{db} gave much better sparsity than other explored wavelets and DCT bases.\\
For the sake of clarity and transparency, it should be mentioned that the image was first transformed to the wavelet domain, and then multiplied with the sensing matrix. In the hardware implementation however, we obtain $\phi x$ (i.e. apply the compression on the original image) as $y$, we then find an $x$ with minimum $l_1$ norm, subject to $y = \phi \psi'x$.  Both the methods produce the exact same result. The former was relatively faster and slightly more convenient, thus was chosen as the preferred way of implementation.\\
Now comes the critical question of determining the number of compressed samples to be considered for inference. I simply began reconstructing random ‘good’ images for different values of $M$ starting with $M = 1000$ (28.6\% compression) and decreasing it with each step. It was realised that even for $M = 500$ (14.3\% compression) the images were somewhat readable. When $M$ was brought down to 200 (5.7\%), it was seen that the text can no longer be read reliably. Thus the number of samples were limited to a maximum of 200. The inference results were then calculated for even lower values. As they began rendering incredible performance on a sample set, 50 (1.4\%), 20 (0.57\%) and 10 (0.28\%) were additionally chosen for experimentation. The reconstruction result for various values of $M$ can be seen in Fig. \ref{fig:recresults}.

\section{Data Decryption}
\label{sec:dec}
The decryption part is sort of an adversary to the objective of data privacy. In general, the sensing matrix need never exist mathematically if the user decides not to note it. But for the sake of assurance they may want to decipher the acquired data. This would require the exact knowledge of the key \textit{i.e.} This encryption itself should be a sufficient level of security for most applications. However, in the event of an unsolicited disclosure of the key (or simply when the system is setup by an external agent who we do not trust to have deleted the key), the user might want to be assured that such decryption efforts are futile.

Since we are considering the worst case scenario, the best possible reconstruction method should be chosen regardless of the computational complexity. It is also assumed that the compressed samples have zero error in measurement. The objective is to ascertain that even with the best resources and maximum information at hand, the reconstruction will render no meaningful result. Thus Basis Pursuit was chosen as the sole reconstruction method.

The convex optimisation package CVX \cite{cvx,cvx2} was used to solve the Basis Pursuit problem. A typical reconstruction with $M= 1000$ took about 200 seconds while it took \textless  2 seconds for $M= 10$.

\section{Classification Methods and Results Analysis}
\label{sec:results}
\begin{figure}
    \centering
    \includegraphics[scale = 0.4]{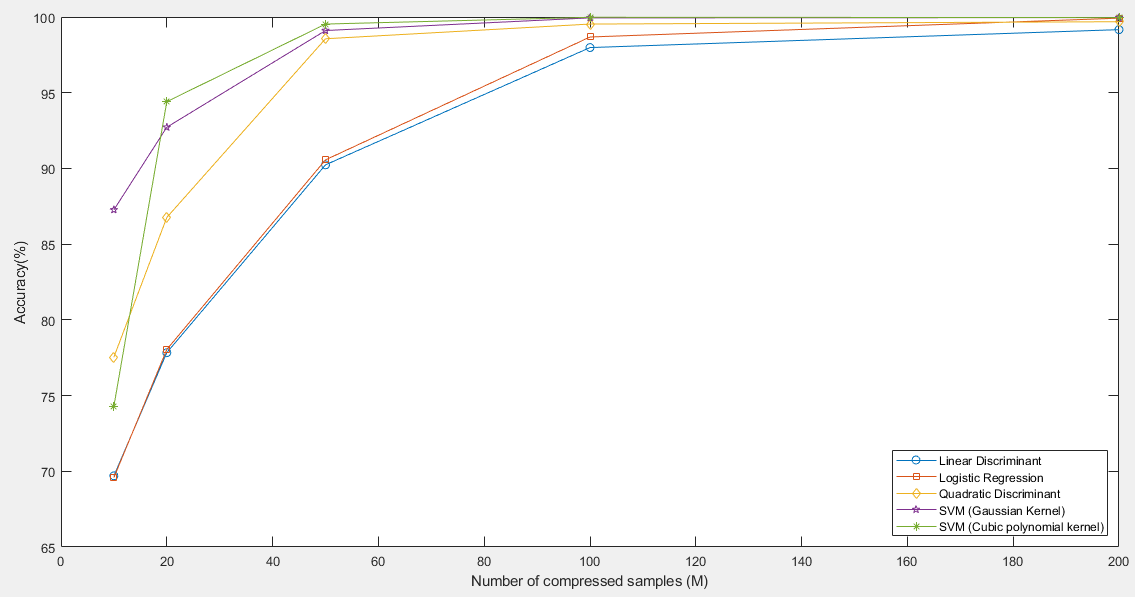}
    \caption{Performance of various classification models versus the number of compressed samples. Refer Table \ref{tab:table} for the precise values.}
    \label{fig:graph}
\end{figure}
Once the maximum number of compressed samples was established, the dataset was bifurcated into training and testing sets. 15000 images from each label were chosen for training while 2576 were chosen for testing from each label. The $30000 \times M$ matrix of training data was augmented with a column of categorical label vector (0 for good, 1 for bad).
Several different classification models were tested with various different parameter values. 5-fold validation was used to compare the results of the different models. It was realised that Support Vector Machines were the best performing among all. The choice of the kernel function and scale significantly affected the performance. In general, cubic kernel function gave the best performance. The kernel scale varied largely for different values of $M$. The performance shown in Fig. \ref{fig:graph} and Table \ref{tab:table} can be considered to have the most optimised parameter values.
$k$-Nearest Neighbour did not perform as good as other models. Performance by Decision trees and Ensemble classifiers like Boosted trees, bagged trees, subspace discriminant, performed relatively sub par as well. Simple classifiers like Linear and Quadratic discriminant, Logistic Regression, performed exceptionally. Though the difference was not huge, these later classifiers provide a more intuitive understanding of how the different categories are separated in the $M-$dimensional hyperspace.

\begin{table}[]
    \centering
    \begin{tabular}{| p{1.5cm} | p{1.5cm} | p{1.5cm} | p{1.5cm} | p{1.5cm} | p{1.5cm} |}
        \hline
         $M$ & Gaussian SVM & Cubic SVM & QD & LR & LD \\
         \hline
         200 & 99.9612 & 99.9612 & 99.6894 & 99.9224 & 99.1654\\
         100 & 99.9418 & 99.9806 & 99.5342 & 98.6801 & 97.9814\\
         50 & 99.1071 & 99.5342 & 98.5637 & 90.5668 & 90.2368\\
         20 & 92.7213 & 94.4099 & 86.7624 & 78.0474 & 77.8339\\
         10 & 87.2671 & 74.3012 & 77.5427 & 69.5846 & 69.7011\\
         \hline
    \end{tabular}
    \caption{Performance of different classifiers. QD: Quadratic Discrimnator, LD: Linear Discriminator, LR: Logical Regression. The accuracy's are in \%. Figure \ref{fig:graph} is a graphical representation of the table.}
    \label{tab:table}
\end{table}

Looking at Table \ref{tab:table}, the general order of performance would be Cubic SVM \textgreater Gaussian SVM \textgreater Quadratic discriminant \textgreater Logistic Regression \textgreater Linear discriminant. Random subsets of misclassified images were analysed for all these models. In general the mix contained all types of ‘bad’ images in about equal proportion. Thus we can be assured that the classifier did not fixate on any particular measurement. However there are two observations that need to be stated.
The most pertaining observation was the consistent asymmetry between the types of misclassification errors – The bad images were classified as good more often than vice versa. The chance exception was in the case of the quadratic discriminant for $M = 200$ wherein all the erroneous classifications were for ‘good’ images which is an outstanding aberration. Another odd consistency was regarding the characters in the misclassified images. Despite all being equal in frequency, the character ‘W’ happened to be in most of them, regardless of the label. Other frequent characters were ‘N’, ‘M’ and ‘X’. It can be observed that these characters have the most prominent diagonal segments. But that is merely stating correlations without establishing causality. A concrete reason for neither of these oddities could be established. I heartedly welcome any analysis the reader would like to share which could shed more light on these observations.

\section{Future Work}
\label{sec:futwork}

As and when such a system is designed, using it in detecting print errors would be the least of its use.
While advances in artificial intelligence has increased the intelligence quotient of our computers, incrementing Emotional intelligence is still a humongous and very coveted task. Although we now have a plethora of AI architectures, they all have limited scope of implementation where user/data privacy is a concern. If we can establish that human emotions can be inferred from digital portraits with a remarkable accuracy via such limited and incoherent measurement that the original image can never be reconstructed to reveal the user identity, it would be landmark in the field of emotion recognition. Having said that, the face database would be much less cooperative than the one used here as faces are much more diverse and have relatively small features that help in accurate detection. This idea can be extended to any field where data privacy is as important as accurate inference, if not more.

Many crucial analysis were left out in this work for the sake of accentuating the proof of concept. The SPC modelled Poisson noise in its measurements and pursued the reconstruction accordingly. Such an addition might affect accuracy or privacy either favourably or otherwise.
Deep learning empowered reconstruction algorithms like Reconnet \cite{reconnet}  Stacked Denoising Auto-encoders (SDAs) \cite{sda} have lately shown remarkable efficacy in reconstructing images from very few compressed samples (as low as 1\%). Testing the data privacy against such reconstruction algorithms should be an imperative course in the line of its advancement. On the other hand, deep neural networks are known to be much \say{deeper} in understanding the classifier of many datasets which are otherwise difficult to extricate. Such advancements should invigorate both objectives of this paper.\\
There are many successful variants of compressive sensing which make use of the structure of data to provide even better reconstructions. Among the most pertaining ones would be block sparsity based reconstruction \cite{block}. Since our data is clearly block sparse in spatial as well as transformed domain, such an algorithm is likely to deliver better results than the basic basis pursuit used here.

\section{Conclusion}

Our mantra in this paper has been to \textit{\say{maximise inference while minimising acquisition}}. We took the print error detection problem in reprography – where the sensitivity of the print data forbids the digitisation – to demonstrate that using \textit{compressive classification}, we can achieve remarkable accuracy in feature detection while ensuring that the data is practically unaccessed. Even at compression ratio as low as 0.3\% and 0.6\%, the results were very satisfactory. We tested multiple models to achieve the desired classification rate. Kernel SVMs (Cubic and Gaussian specifically) turned out to be most efficient in general. It was observed that even simple linear discriminants provided astounding accuracy when the compression ratio was relatively higher, but not high enough to provide meaningful reconstruction.

\bibliographystyle{splncs03_unsrt}
\bibliography{references}

\begin{thebibliography}{10}
\providecommand{\url}[1]{\texttt{#1}}
\providecommand{\urlprefix}{URL }

\bibitem{candromtao}
{Candes}, E.J., {Romberg}, J., {Tao}, T.: Robust uncertainty principles: exact
  signal reconstruction from highly incomplete frequency information. IEEE
  Transactions on Information Theory  52(2),  489--509 (Feb 2006)

\bibitem{donoho}
{Donoho}, D.L.: Compressed sensing. IEEE Transactions on Information Theory
  52(4),  1289--1306 (April 2006)

\bibitem{cstut}
{Candes}, E.J., {Wakin}, M.B.: An introduction to compressive sampling. IEEE
  Signal Processing Magazine  25(2),  21--30 (March 2008)

\bibitem{samplingtheorem}
{Vaidyanathan}, P.P.: Generalizations of the sampling theorem: Seven decades
  after nyquist. IEEE Transactions on Circuits and Systems I: Fundamental
  Theory and Applications  48(9),  1094--1109 (Sep 2001)

\bibitem{pareto}
Van Den~Berg, E., Friedlander, M.P.: Probing the pareto frontier for basis
  pursuit solutions. SIAM Journal on Scientific Computing  31(2),  890--912
  (2008)

\bibitem{omp}
{Rebollo-Neira}, L., {Lowe}, D.: Optimized orthogonal matching pursuit
  approach. IEEE Signal Processing Letters  9(4),  137--140 (April 2002)

\bibitem{romp}
Tropp, J.A., Gilbert, A.C.: Signal recovery from random measurements via
  orthogonal matching pursuit. IEEE Transactions on information theory  53(12),
   4655--4666 (2007)

\bibitem{stomp}
Donoho, D.L., Drori, I., Tsaig, Y., Starck, J.L.: Sparse solution of
  underdetermined linear equations by stagewise orthogonal matching pursuit.
  Department of Statistics, Stanford University (2006)

\bibitem{cosamp}
Needell, D., Tropp, J.A.: Cosamp: Iterative signal recovery from incomplete and
  inaccurate samples. Applied and computational harmonic analysis  26(3),
  301--321 (2009)

\bibitem{spc}
{Duarte}, M.F., {Davenport}, M.A., {Takhar}, D., {Laska}, J.N., {Sun}, T.,
  {Kelly}, K.F., {Baraniuk}, R.G.: Single-pixel imaging via compressive
  sampling. IEEE Signal Processing Magazine  25(2),  83--91 (March 2008)

\bibitem{smash}
A.~Davenport, M., Duarte, M., B.~Wakin, M., N.~Laska, J., Takhar, D., Kelly,
  K., Baraniuk, R.: The smashed filter for compressive classification and
  target recognition - art. no. 64980h. Proceedings of SPIE  6498 (02 2007)

\bibitem{ppcs}
{Bianchi}, T., {Bioglio}, V., {Magli}, E.: Analysis of one-time random
  projections for privacy preserving compressed sensing. IEEE Transactions on
  Information Forensics and Security  11(2),  313--327 (Feb 2016)

\bibitem{irtrack}
{Braun}, H., {Turaga}, P., {Spanias}, A.: Direct tracking from compressive
  imagers: A proof of concept. In: 2014 IEEE International Conference on
  Acoustics, Speech and Signal Processing (ICASSP). pp. 8139--8142 (May 2014)

\bibitem{ttw}
{Balthasar}, M.R., {Leigsnering}, M., {Zoubir}, A.M.: Compressive
  classification for through-the-wall radar imaging. In: 2012 Proceedings of
  the 20th European Signal Processing Conference (EUSIPCO). pp. 2288--2292 (Aug
  2012)

\bibitem{jrom}
{Amaravati}, A., {Xu}, S., {Romberg}, J., {Raychowdhury}, A.: A 130 nm 165
  nj/frame compressed-domain smashed-filter-based mixed-signal classifier for
  “in-sensor” analytics in smart cameras. IEEE Transactions on Circuits and
  Systems II: Express Briefs  65(3),  296--300 (March 2018)

\bibitem{crypt1}
{Rachlin}, Y., {Baron}, D.: The secrecy of compressed sensing measurements. In:
  2008 46th Annual Allerton Conference on Communication, Control, and
  Computing. pp. 813--817 (Sep 2008)

\bibitem{crypt18}
{Orsdemir}, A., {Altun}, H.O., {Sharma}, G., {Bocko}, M.F.: On the security and
  robustness of encryption via compressed sensing. In: MILCOM 2008 - 2008 IEEE
  Military Communications Conference. pp. 1--7 (Nov 2008)

\bibitem{crypt19}
{Cambareri}, V., {Haboba}, J., {Pareschi}, F., {Rovatti}, H.R., {Setti}, G.,
  {Wong}, K.: A two-class information concealing system based on compressed
  sensing. In: 2013 IEEE International Symposium on Circuits and Systems
  (ISCAS2013). pp. 1356--1359 (May 2013)

\bibitem{crypt20}
{Cambareri}, V., {Mangia}, M., {Pareschi}, F., {Rovatti}, R., {Setti}, G.:
  Low-complexity multiclass encryption by compressed sensing. IEEE Transactions
  on Signal Processing  63(9),  2183--2195 (May 2015)

\bibitem{crypt21}
{Cambareri}, V., {Mangia}, M., {Pareschi}, F., {Rovatti}, R., {Setti}, G.: On
  known-plaintext attacks to a compressed sensing-based encryption: A
  quantitative analysis. IEEE Transactions on Information Forensics and
  Security  10(10),  2182--2195 (Oct 2015)

\bibitem{pprs}
Zhou, S., Lafferty, J., Wasserman, L.: Compressed and privacy-sensitive sparse
  regression. IEEE Transactions on Information Theory  55(2),  846--866 (2009)

\bibitem{matrixmasking}
Duncan, G.T., Pearson, R.W., et~al.: Enhancing access to microdata while
  protecting confidentiality: Prospects for the future. Statistical Science
  6(3),  219--232 (1991)

\bibitem{db}
Daubechies, I.: Ten lectures on wavelets, vol.~61. Siam (1992)

\bibitem{cvx}
Grant, M., Boyd, S.: {CVX}: Matlab software for disciplined convex programming,
  version 2.1. \url{http://cvxr.com/cvx} (Mar 2014)

\bibitem{cvx2}
Grant, M., Boyd, S.: Graph implementations for nonsmooth convex programs. In:
  Blondel, V., Boyd, S., Kimura, H. (eds.) Recent Advances in Learning and
  Control, pp. 95--110. Lecture Notes in Control and Information Sciences,
  Springer-Verlag Limited (2008),
  \url{http://stanford.edu/~boyd/graph_dcp.html}

\bibitem{reconnet}
Kulkarni, K., Lohit, S., Turaga, P., Kerviche, R., Ashok, A.: Reconnet:
  Non-iterative reconstruction of images from compressively sensed
  measurements. In: Proceedings of the IEEE Conference on Computer Vision and
  Pattern Recognition. pp. 449--458 (2016)

\bibitem{sda}
Mousavi, A., Patel, A.B., Baraniuk, R.G.: A deep learning approach to
  structured signal recovery. In: 2015 53rd Annual Allerton Conference on
  Communication, Control, and Computing (Allerton). pp. 1336--1343. IEEE (2015)

\bibitem{block}
Eldar, Y.C., Bolcskei, H.: Block-sparsity: Coherence and efficient recovery.
  In: 2009 IEEE International Conference on Acoustics, Speech and Signal
  Processing. pp. 2885--2888. IEEE (2009)

\end{thebibliography}
\end{document}